\documentclass[aps,prb,twocolumn,amsmath,amssymb,nofootinbib,eqsecnum,superscriptaddress,floatfix]{revtex4}

\usepackage{amsmath}
\usepackage{amssymb}
\usepackage{amsthm}
\usepackage[dvips]{color}
\usepackage{graphicx}% Include figure files
\usepackage{bm} % bold math
\usepackage[hypertex]{hyperref}
\usepackage{ulem}   % to strike things out
\newcommand{\bs}[1]{{\boldsymbol{#1}}}

\begin{document}

\title{ 
Elementary formula for the Hall conductivity of interacting systems
      }

\date{\today}

\author{Titus Neupert} 
\affiliation{
Condensed Matter Theory Group, 
Paul Scherrer Institute, CH-5232 Villigen PSI,
Switzerland
            } 

\author{Luiz Santos} 
\thanks{
Present address:
Perimeter Institute for Theoretical Physics, 
Waterloo, Ontario, Canada N2L 2Y5.
       }
\affiliation{
Department of Physics, 
Harvard University, 
17 Oxford Street, 
Cambridge, Massachusetts 02138, USA
            } 

\author{Claudio Chamon} 
\affiliation{
Physics Department, 
Boston University, 
Boston, Massachusetts 02215, USA
            } 

\author{Christopher Mudry} 
\affiliation{
Condensed Matter Theory Group, 
Paul Scherrer Institute, CH-5232 Villigen PSI,
Switzerland
            } 

\begin{abstract}
A formula for the Hall conductivity of interacting electrons is given
under the assumption that the ground state manifold is
$N^{\ }_{\mathrm{gs}}$-fold degenerate and discrete translation symmetry is
neither explicitly nor spontaneously broken. 
\end{abstract}

\maketitle

\section{
Introduction and results
        }
\label{sec: Introduction and results}

The Hall conductivity in the 
integer (IQHE)
and fractional (FQHE)
quantum Hall effect
is observed to be quantized.%
~\cite{FQHE}  
This quantization
can be explained by expressing the Hall conductivity 
$\sigma^{\ }_{\mathrm{H}}$
in terms of a topological invariant
$C$ that takes integer values
through the relation
\begin{equation}
\sigma^{\ }_{\mathrm{H}}=
\frac{e^{2}}{h}\,
\nu\,
C
\end{equation}
where $\nu$ is the filling fraction of the single-particle
Landau levels.%
~\cite{Laughlin81,Thouless82,Avron83,Niu84,Niu85}
For example,
the relation between the quantized number $C$
and the \textit{non-degenerate} many-body ground state wave functions
$|\Psi(\phi,\varphi)\rangle$ 
obeying twisted boundary conditions parametrized by 
the pair of angles $0\leq\phi,\varphi\leq2\pi$
for a gas of electrons confined in 
two-dimensional (position) space
and subjected to a uniform magnetic field is
\begin{equation}
C=
-
\frac{\mathrm{i}}{2\pi}
\int\limits_{0}^{2\pi}\mathrm{d}\phi
\int\limits_{0}^{2\pi}\mathrm{d}\varphi
\left[
\left\langle
\frac{\partial\Psi}{\partial\phi}
\right|
\left.
\frac{\partial\Psi}{\partial\varphi}
\right\rangle
-
\left\langle
\frac{\partial\Psi}{\partial\varphi}
\right|
\left.
\frac{\partial\Psi}{\partial\phi}
\right\rangle
\right],
\end{equation}
when the cyclotron energy dominates over the characteristic
electron-electron interactions while the characteristic
spatial variations of any one-body potential experienced
by the electrons
are much longer than the magnetic length.%
~\cite{Niu84,Niu85}

\begin{subequations}
In this paper, we derive a formula for the Hall conductivity
that applies to electrons moving in $d$-dimensional space
as long as 
(i) there exists a gap between the 
$N^{\ }_{\mathrm{gs}}$-fold degenerate ground state manifold
and
(ii) discrete translation invariance holds.%
~\cite{footnote: meaning translation inavriance}
We show that the
Hall conductivity averaged over the degenerate ground states
is given by Eq.~(\ref{eq: main rep bar sigma Bloch}),
which reduces to 
\begin{eqnarray}
&&\bar{\sigma}^{\ }_{\mathrm{H}}=
\frac{e^2}{h}
\int\limits_{\Omega}
\mathrm{d}^2\bs{k}\;
F(\bs{k})\;
\bar{\mathrm{n}}(\bs{k}),
\label{eq: main result intro}
\\
&&
F(\bs{k}):=
\mathrm{i}
\langle
\partial^{\ }_{k^{\ }_2}\chi(\bs{k})
|
\partial^{\ }_{k^{\ }_1}\chi(\bs{k})
\rangle
-(1\leftrightarrow2),
\end{eqnarray}
\end{subequations}
in two-dimensional space 
if restricted to the case when the 
many-body ground states
are exclusively built out of a single Bloch band with Bloch states $|\chi(\bs{k})\rangle$
and the single-particle Berry curvature $F(\bs{k})$. 
All the many-body correlations in 
Eq.~(\ref{eq: main result intro})
are encoded by 
$\bar{\mathrm{n}}(\bs{k})$ 
defined in Eq.~(\ref{eq: def many-body nab(k)}),
i.e., the expectation value 
of the occupation number operator of the Bloch momentum 
$\bs{k}$ from the Brillouin zone with volume $\Omega$  
averaged over the 
$N^{\ }_{\mathrm{gs}}$-fold
degenerate many-body ground states.

Equation~(\ref{eq: main result intro}) 
reproduces the following results.

\textit{Cases of the IQHE or the Chern band insulator.}
When $N^{\ }_{\mathrm{gs}}=1$ and the 
many-body ground state is the Slater determinant
made of all available Bloch states in the band,
$\bar{\mathrm{n}}(\bs{k})=1$ for all $\bs{k}\in\Omega$
and
\begin{equation}
\bar{\sigma}^{\ }_{\mathrm{H}}=
\frac{e^2}{h}
\int\limits_{\Omega}
\mathrm{d}^2\bs{k}\;
F(\bs{k})=:
\frac{e^2}{h}\;C
\label{eq:formula_Bn_IQHE}
\end{equation}
with $C$ the Chern number of the band 
(in particular, the Berry curvature is the constant
$F(\bs{k})=1/\Omega$
and the Chern number is $C=1$ 
for the lowest Landau level).

\textit{Case of the FQHE.}
When $N^{\ }_{\mathrm{gs}}>1$ is the degeneracy expected from
a hierarchical ground state in the FQHE at 
the partial filling $\nu$
of the lowest Landau level, the Berry curvature 
is the constant $F(\bs{k})=C/\Omega$
with $C=1$ the Chern number of the lowest Landau level and
\begin{equation}
\bar{\sigma}^{\ }_{\mathrm{H}}=
\frac{e^2}{h}
\int\limits_{\Omega}
\frac{\mathrm{d}^2\bs{k}}{\Omega}\;
\bar{\mathrm{n}}(\bs{k})
=:
\frac{e^2}{h}\;\nu.
\label{eq:formula_Bn_FQHE}
\end{equation}

\textit{Case of the anomalous Hall effect.}
When $N^{\ }_{\mathrm{gs}}=1$ and the 
many-body ground state is a Fermi liquid,
we can interpret Eq.~(\ref{eq: main result intro}) 
as the anomalous Hall conductivity
assuming the order of limits by which
the gap has been taken to zero after all other limits
have been taken.
The anomalous Hall conductivity has been derived 
in the noninteracting limit,%
~\cite{Karpus54,Ong06,AHE}
but its derivation allows for interactions in this paper.

Equation~(\ref{eq: main result intro}) 
predicts the following results for fractional Chern insulators.%
~\cite{footnote: def FCI,Neupert11,Sheng11,Wang11a,Regnault11a,Bernevig11,Wang12a,Bernevig12a,Bernevig12b,Wang12b,Liu12}
 
\begin{enumerate}
\item[(1)] 
The integral over the Brillouin zone of
$F(\bs{k})\times\bar{\mathrm{n}}(\bs{k})$
equals a rational number $p/q$,
since Laughlin's gauge argument for quantization 
then applies.%
~\cite{Laughlin81}
\item[(2)] 
The integral over the Brillouin zone of
$F(\bs{k})\times\bar{\mathrm{n}}(\bs{k})$
obeys a sum rule, for one cannot change the rational number 
$p/q$ continuously.

\item[(3)] 
The Hall conductivity $\bar{\sigma}^{\ }_{\mathrm{H}}$
does not need to be equal to the filling fraction $\nu$ 
obtained by integrating $\bar{\mathrm{n}}(\bs{k})/\Omega$ 
over $\Omega$ whenever the Berry curvature
$F(\bs{k})$ 
is not uniform over the Brillouin zone. 
In this case, 
if $F^{\ }_{\mathrm{min}}$ 
and $F^{\ }_{\mathrm{max}}$ are
the minimum and maximum of $F(\bs{k})$ 
over the Brillouin zone, respectively, then 
\begin{equation}
\qquad\quad
\frac{e^2}{h}\,\nu\,
\left(\Omega\,F^{\ }_{\mathrm{min}}\right)
\le
\bar{\sigma}^{\ }_{\mathrm{H}}
\le
\frac{e^2}{h}\,\nu\,
\left(\Omega\,F^{\ }_{\mathrm{max}}\right).
\label{eq:formula_Bn_FCI}
\end{equation}
\end{enumerate}

The reminder of the paper is devoted to deriving these results.

\section{
Definitions
        }
\label{sec: Definitions} 

We consider $N^{\ }_{\mathrm{e}}$ electrons 
confined to a box of volume $V$
in position space $\mathbb{R}^{d}$ where $d=1,2,\cdots$
whose dynamics is governed
by the conserved many-body Hamiltonian $H^{\ }_{0}$ 
acting on the many-body Hilbert space
$\mathfrak{F}^{\ }_{N^{\ }_{\mathrm{e}}}$.
Boundary conditions have been imposed such that
$H^{\ }_{0}$ has a countable number of eigenstates.
The spectral decomposition of $H^{\ }_{0}$
is written
\begin{subequations}
\begin{equation}
H^{\ }_{0}=
E^{\ }_{\mathrm{gs}}
\sum_{n=1}^{N^{\ }_{\mathrm{gs}}}
|n\rangle\langle n|
+
\sum_{m=N^{\ }_{\mathrm{gs}}+1}^{\infty}
E^{\ }_{m}
|m\rangle\langle m|.
\end{equation}
The ground state eigenenergy $E^{\ }_{\mathrm{gs}}$
is assumed to be $N^{\ }_{\mathrm{gs}}$-fold degenerate.
We reserve the integer-valued index $n=1,\cdots,N^{\ }_{\mathrm{gs}}$ 
for its $N^{\ }_{\mathrm{gs}}$ linearly independent ground states. 
A gap separating $E^{\ }_{\mathrm{gs}}$
from the eigenenergies $E^{\ }_{m}$ 
of all excited states 
is assumed,
whereby we reserve
the integer-valued index 
$m=N^{\ }_{\mathrm{gs}}+1,\cdots$
for the excited states.
The eigenstates of $H^{\ }_{0}$
are normalized according to 
\begin{equation}
\langle n|n'\rangle=\delta^{\ }_{n,n'},
\quad
\langle m|m'\rangle=\delta^{\ }_{m,m'},
\quad
\langle m|n\rangle=0.
\label{eq: normalization many-body states n and m}
\end{equation}
\end{subequations}
This means that the
scalar product of any two states from the Hilbert space
$\mathfrak{F}^{\ }_{N^{\ }_{\mathrm{e}}}$
is dimensionless.
Finally, the total electric charge is assumed conserved
by $H^{\ }_{0}$ and this symmetry is not spontaneously broken.

Because of the boundary conditions,
the set of center-of-mass momenta is countable, i.e.,
we can resolve the identity according to
\begin{equation}
1=
\sum_{\bs{Q}}
P^{\ }_{\bs{Q}},
\qquad
P^{\ }_{\bs{Q}}\,
P^{\ }_{\bs{Q}'}=
\delta^{\ }_{\bs{Q},\bs{Q}'}
P^{\ }_{\bs{Q}},
\label{eq: resolution 1 in terms PQ and ortho PQ's}
\end{equation}
where $\bs{Q}\in\mathbb{R}^{d}$ 
denotes a center-of-mass momentum
and $P^{\ }_{\bs{Q}}$ denotes the projector onto states
with this center-of-mass momentum.

Finally, we shall assume that $H^{\ }_{0}$ 
does not break discrete translational invariance either,%
~\cite{footnote: meaning translation inavriance} 
i.e., 
\begin{equation}
[H^{\ }_{0},P^{\ }_{\bs{Q}}]=0
\end{equation}
for any center-of-mass momentum $\bs{Q}$, nor is
discrete translational invariance spontaneously broken.

We work in units where the electric charge $e$, 
the Planck constant $\hbar$, and the speed of light $c$ are all unity.
Therefore, only dimensions of energy appear in the theory.

\section{
Linear response
        }
\label{sec: Linear response}

We want to find the linear response of one ground state $|n\rangle$
of  the many-body Hamiltonian $H^{\ }_{0}$ to adiabatically switching 
on a spatially homogeneous and static electric field $\bs{\mathcal{E}}$. 
The coupling to $\bs{\mathcal{E}}$ is described by the Hamiltonian
\begin{equation}
H:=
H^{\ }_{0}
+
H^{\ }_{1}(t),
\qquad
H^{\ }_{1}(t):=
-
\bm{X}
\cdot
\bm{\mathcal{E}}\,
e^{\eta\,t},
\label{eq: def H in S picture}
\end{equation}
where  $\bs{X}\equiv(X^{\ }_{i})$ 
is the many-body position operator in the Schr\"odinger picture,%
~\cite{footnote: def X,Neupert12b}
$i=1,\cdots,d$ labels the spatial coordinates, and
$\eta$ is a small positive number that implements
the adiabatic turn-on of
$\bm{\mathcal{E}}$. 

We would like to make explicit the existence of a conserved 
electronic current that couples to the applied static 
and uniform electric field
$\bm{\mathcal{E}}$. To this end, observe that we can always write
\begin{subequations}
\label{eq: def E in terms A of t}
\begin{equation}
\bm{\mathcal{E}}=
-
\left(\partial^{\ }_{t}\bm{\mathcal{A}}\right)(t),
\qquad
\bm{\mathcal{A}}(t):=
-t\,\bm{\mathcal{E}}
\label{eq: def E in terms A of t a}.
\end{equation}
Insertion of Eq.~(\ref{eq: def E in terms A of t a})
into Eq.~(\ref{eq: def H in S picture})
gives
\begin{equation}
H=
H^{\ }_{0}
-
\bm{J}
\cdot
\bm{\mathcal{A}}(t)\,
e^{\eta\,t}
+
\partial^{\ }_{t}
\left[
\bm{X}
\cdot
\bm{\mathcal{A}}(t)\,
e^{\eta\,t}
\right]
\label{eq: def E in terms A of t b}
\end{equation}
with the current%
~\cite{footnote: dimensions}
\begin{equation}
\bm{J}:=
\mathrm{i}
\left[
H^{\ }_{0},
\bm{X}
\right]
\equiv
\partial^{\ }_{t}\bm{X}.
\label{eq: def H in S picture c}
\end{equation}
\end{subequations}
In effect, we have performed the time-dependent gauge
transformation 
\begin{equation}
\begin{split}
&
\widetilde{H}(t):=
\widetilde{H}^{\ }_{0}
-
\widetilde{\bm{J}}
\cdot
\bm{\mathcal{A}}(t)\,
e^{\eta\,t},
\\
&
\widetilde{H}^{\ }_{0}(t):=
e^{
\mathrm{+i}
\bm{X}
\cdot
\bm{\mathcal{A}}(t)\,
e^{\eta\,t}
  }
H^{\ }_{0}\,
e^{
\mathrm{-i}
\bm{X}
\cdot
\bm{\mathcal{A}}(t)\,
e^{\eta\,t}
  },
\\
&
\widetilde{\bs{J}}(t):=
\mathrm{i}
\left[
\widetilde{H}^{\ }_{0}(t),
\bm{X}
\right].
\end{split}
\end{equation}

In anticipation of linear response theory,
we are going to use the interaction picture
that endows any operator $O^{\ }$
in the Schr\"odinger picture with the time dependence
\begin{equation}
O^{\mathrm{I}}(t):=
e^{+\mathrm{i}\,H^{\ }_{0}\,t}\,
O^{\ }\, 
e^{-\mathrm{i}\,H^{\ }_{0}\,t}.
\label{eq: def interation picture}
\end{equation}
We do the same with the instantaneous
ground states $|\widetilde{n}\rangle^{\ }_{t}$
and the instantaneous excited states 
$|\widetilde{m}\rangle^{\ }_{t}$
of
$\widetilde{H}^{\ }_{0}(t)$.
In linear response theory, 
we approximate the time evolution of 
any state from $\mathfrak{F}^{\ }_{N^{\ }_{\mathrm{e}}}$ in the interaction picture
by linearizing in the perturbation 
\begin{equation}
H^{\mathrm{I}}_{1}(t)=
-
\bm{J}^{\mathrm{I}}
\cdot
\bm{\mathcal{A}}(t)\,
e^{\eta\,t}
+
\partial^{\ }_{t}
\left[
\bm{X}^{\mathrm{I}}
\cdot
\bm{\mathcal{A}}(t)\,
e^{\eta\,t}
\right].
\end{equation}
For example, for one of the degenerate ground states,
\begin{equation}
|n(t)\rangle^{\mathrm{I}}:=
\left(
1
-
\mathrm{i}
\int\limits_{-\infty}^{t}
\mathrm{d} t'\, 
H^{\mathrm{I}}_{1}(t')
\right)
|n\rangle
+
\cdots\,.
\label{eq: example interaction picture n}
\end{equation}
Hence,
if we rule out any level crossing
[see Eq.~(\ref{eq: justification degenerate perturbation theory})
and Ref.~\onlinecite{Niu85}], 
then
\begin{equation}
\begin{split}
{}^{\ }_{t}
\big\langle\widetilde{n}\big|
\bm{\widetilde{J}}(t)\,
\big|\widetilde{n}\big\rangle
{}^{\ }_{t}
=&\,
{}^{\mathrm{I}}\!
\left\langle n(t)\left|
\bm{J}^{\mathrm{I}}(t)
\right|n(t)\right\rangle\!
{}^{\mathrm{I}}
\\
=&\,
-\mathrm{i}\,
\int\limits_{-\infty}^{t}\mathrm{d} t^{\ }_1
\left\langle n\left|
\left[
\bm{J}^{\mathrm{I}}(t),
H^{\mathrm{I}}_{1}(t^{\ }_{1})
\right]
\right|n\right\rangle
\end{split}
\end{equation}
to linear order in the time-dependent perturbation and after
making use of the identity
[following Eq.~(\ref{eq: def H in S picture c})]
\begin{equation}
\left\langle n\left|\bm{J}^{\mathrm{I}}(t)\right|n\right\rangle=
\left\langle n\left|\bm{J}\right|n\right\rangle=
0
\end{equation}
that disposes of the lowest-order contribution to the expansion.
With some intermediary steps involving the integral representation
$
\bm{J}^{\mathrm{I}}(t^{\ }_{1})=
\frac{\mathrm{d}}{\mathrm{d}t^{\ }_{1}}
\int\limits_{-\infty}^{t^{\ }_{1}}
\mathrm{d}t^{\ }_{2}\,
\bm{J}^{\mathrm{I}}(t^{\ }_{2})
$
and the use of partial integration,
we arrive for any given $i=1,\cdots,d$
at the leading-order estimate
(with summation convention over the repeated index $j=1,\cdots,d$)
\begin{equation}
\begin{split}
{}^{\ }_{t}
\big\langle\widetilde{n}\big|
\widetilde{J}^{\ }_{i}(t)\,
\big|\widetilde{n}\big\rangle
{}^{\ }_{t}
=&\,
\mathrm{i}\,
\mathcal{E}^{\ }_{j}
\int\limits_{-\infty}^{t}\!\mathrm{d}t^{\ }_{1}\!
\int\limits_{-\infty}^{t^{\ }_{1}}\!\mathrm{d}t^{\ }_{2}
\left\langle n\left|
\left[
J^{\mathrm{I}}_{i}(t), 
J^{\mathrm{I}}_{j}(t^{\ }_{2})
\right] 
\right|n\right\rangle.
\end{split}
\end{equation}
Finally,
insertion of the resolution of the identity in 
$\mathfrak{F}^{\ }_{N^{\ }_{\mathrm{e}}}$
\begin{equation}
1=
\sum_{n}|n\rangle\langle n|
+
\sum_{m}|m\rangle\langle m|
\end{equation}
delivers
\begin{equation}
\begin{split}
\vphantom{{A}^{A}}_{t}
\big\langle\widetilde{n}\big|
\widetilde{J}^{\ }_{i}(t)\,
\big|\widetilde{n}\big\rangle
\vphantom{{A}^{A}}_{t}
=&\,
-
\mathrm{i}\,
\mathcal{E}^{\ }_{j}
\sum_m
\frac{
\langle n|J^{\ }_{i}|m\rangle 
\langle m|J^{\ }_{j}|n\rangle
-
(i\leftrightarrow j)
     }
     {
\left(E^{\ }_{m}-E^{\ }_{\mathrm{gs}}\right)^{2}
     }.
\end{split}
\end{equation}
Here, we used that 
\begin{equation}
\langle n|\bm{J}|n'\rangle=0
\label{eq: justification degenerate perturbation theory}
\end{equation}
for any pair $n,n'=1,\cdots,N^{\ }_{\mathrm{gs}}$
in view of Eq.~(\ref{eq: def H in S picture c}).
Equation~(\ref{eq: justification degenerate perturbation theory})
prevents level crossing among the
degenerate ground states. For any $i,j=1,\cdots,d$,
we introduce the conductivity tensor 
$\sigma^{(n)}_{ij}$
through
\begin{equation}
\vphantom{{A}^{A}}_{t}
\big\langle\widetilde{n}\big|
\widetilde{J}^{\ }_{i}(t)\,
\big|\widetilde{n}\big\rangle
\vphantom{{A}^{A}}_{t}
=:
2\pi\,V\,
\sigma^{(n)}_{ij}\,
\mathcal{E}^{\ }_{j}.
\label{eq: def conductivity conventional a}
\end{equation}
For any pair $i\neq j=1,\cdots,d$,
the volume $V$ on the right-hand side guarantees that 
the Hall conductivity tensor 
\begin{equation}
\sigma^{(n)}_{ij}=
-
\frac{\mathrm{i}}{2\pi\,V}\,
\sum_{m}
\frac{
\langle n|J^{\ }_{i}|m\rangle 
\langle m|J^{\ }_{j}|n\rangle
-
(i\leftrightarrow j)
     }
     {
\left(E^{\ }_{m}-E^{\ }_{\mathrm{gs}}\right)^{2}
     }
\label{eq: def conductivity conventional b}
\end{equation}
is (i) intensive despite the fact that the left-hand side scales 
with the system size and (ii) dimensionless if $d=2$.
Equation~(\ref{eq: def conductivity conventional b})
is the conventional representation of the Hall conductivity tensor.
The right-hand side is an infinite series which is presumed
convergent because of the energy denominators.

%The right-hand side is often evaluated numerically with
%the help of a regularization that enforces a finite-dimensional
%Hilbert space combined with an exact diagonalization of the interacting 
%Hamiltonian that delivers the exact many-body eigenstates 
%and eigenvalues.

For our purpose, it is, however, more useful
to trade the current for the position operator
in the matrix elements of the right-hand side of Eq.%
~(\ref{eq: def conductivity conventional b})
with the help of Eq.~(\ref{eq: def H in S picture c}) 
and, in turn, dispose of the energy denominator
\begin{equation}
\sigma^{(n)}_{ij}=
-
\frac{
\mathrm{i}
     }
     {
2\pi\, V
     }
\sum_{m}
\left[
\langle n|
X^{\ }_{i} 
|m\rangle 
\langle m|X^{\ }_{j}|n\rangle
-
(i\leftrightarrow j)
\right].
\label{eq: dangerous rep sigma}
\end{equation}
However, we must be very careful with the interpretation of
the matrix element
$\langle m|X^{\ }_{j}|n\rangle$.
In the thermodynamic limit 
$N^{\ }_{\mathrm{e}},V\to\infty$
holding the electronic density $N^{\ }_{\mathrm{e}}/V$ fixed,
the position operator $X^{\ }_{j}$ is unbounded,
its expectation value in any momentum eigenstate
is ill-defined, and so is its trace over the Hilbert space.  
Hence, the representation~(\ref{eq: dangerous rep sigma})
of the Hall conductivity tensor is done in terms of
the difference of two series, each of which is divergent
in the thermodynamic limit $N^{\ }_{\mathrm{e}},V\to\infty$
holding the electronic density $N^{\ }_{\mathrm{e}}/V$ fixed.
Our goal is to properly regularize the subtraction
of two infinities on the right-hand side 
by means of formal algebraic manipulations.

\section{
Algebraic regularization
        }
\label{sec: Algebraic regularization}

For any $i\neq j=1,\cdots,d$,
instead of the Hall conductivity of a single degenerate ground state, 
we define the average 
\begin{equation}
\bar{\sigma}^{\ }_{ij}:=
\frac{1}{N^{\ }_{\mathrm{gs}}}\,
\sum_{n}
\sigma^{(n)}_{ij}.
\label{eq: def mean sigmaij}
\end{equation}
Because of the normalization%
~(\ref{eq: normalization many-body states n and m}),
we can introduce the projector on the ground states
\begin{subequations}
\begin{equation}
P^{\ }_{\mathrm{g}}:=
\sum_{n}  |n\rangle  \langle n|
\label{eq: def Pg}
\end{equation}
and the projector on the excited states
\begin{equation}
P^{\ }_{\mathrm{e}}:=
\sum_m  |m\rangle  \langle m|.
\label{eq: def Pe}
\end{equation}
Evidently, the normalization%
~(\ref{eq: normalization many-body states n and m}) 
implies that
\begin{equation}
\begin{split}
&
P^{2 }_{\mathrm{g}}=P^{\ }_{\mathrm{g}}=P^{\dag}_{\mathrm{g}},
\qquad
P^{2 }_{\mathrm{e}}=P^{\ }_{\mathrm{e}}=P^{\dag}_{\mathrm{e}},
\\
&
P^{\ }_{\mathrm{g}}\,P^{\ }_{\mathrm{e}}=
P^{\ }_{\mathrm{e}}\,P^{\ }_{\mathrm{g}}=0,
\qquad
P^{\ }_{\mathrm{e}}+P^{\ }_{\mathrm{g}}=
1.
\end{split}
\end{equation}
\end{subequations}
In terms of the projectors 
(\ref{eq: def Pg})
and 
(\ref{eq: def Pe}),
Eq.~(\ref{eq: def mean sigmaij}) 
becomes
\begin{equation}
\bar{\sigma}^{\ }_{ij}
=\,
-\frac{\mathrm{i}}{2\pi\,V\,N^{\ }_{\mathrm{gs}}}\,
\mathrm{Tr}\,
\left[
P^{\ }_{\mathrm{g}}
X^{\ }_{i} 
P^{2}_{\mathrm{e}}
X^{\ }_{j}
P^{\ }_{\mathrm{g}}
-
(i\leftrightarrow j)
\right]
\label{eq: cond as trace over X}
\end{equation}
where Tr denotes the trace over the Hilbert space 
$\mathfrak{F}^{\ }_{N^{\ }_{\mathrm{e}}}$.

The discrete translational invariance of the Hamiltonian $H^{\ }_{0}$
guarantees that it commutes with $P^{\ }_{\bs{Q}}$ 
for any center-of-mass momentum $\bs{Q}$,
$
P^{\ }_{\bs{Q}}\,
H^{\ }_{0}=
H^{\ }_{0}\,
P^{\ }_{\bs{Q}}.
$
Correspondingly,
any eigenstate of $H^{\ }_{0}$
can be labeled by a center-of-mass momentum and
\begin{equation}
P^{\ }_{\bs{Q}}\,P^{\ }_{\mathrm{g}}=
P^{\ }_{\mathrm{g}}\,P^{\ }_{\bs{Q}},
\qquad
P^{\ }_{\bs{Q}}\,P^{\ }_{\mathrm{e}}=
P^{\ }_{\mathrm{e}}\,P^{\ }_{\bs{Q}},
\qquad \forall \bs{Q}.
\end{equation}
Since we ruled out spontaneous symmetry breaking
of discrete translational invariance by assumption,
any ground state must 
be an eigenstate of the translation operator.%
~\cite{footnote: symm breaking}
Indeed, if two ground states differ in their center-of-mass momentum,
then not all linear superpositions of them 
are eigenstates of the translation operator.
We conclude that the absence of spontaneous breaking of discrete
translational invariance implies that all states in the ground state manifold
have the same center-of-mass momentum $\bs{Q}^{\ }_0$. If so,
\begin{equation}
P^{\ }_{\bs{Q}}\,P^{\ }_{\mathrm{g}}=
P^{\ }_{\mathrm{g}}\,P^{\ }_{\bs{Q}}=0,
\qquad
\forall
\bs{Q}\neq\bs{Q}^{\ }_0,
\label{eq: PgPQ=0}
\end{equation}
so that, combined with the application
\begin{equation}
P^{\ }_{\bs{Q}}=
P^{\ }_{\bs{Q}}\,
\left(
P^{\ }_{\mathrm{g}}
+
P^{\ }_{\mathrm{e}}
\right)=
\left(
P^{\ }_{\mathrm{g}}
+
P^{\ }_{\mathrm{e}}
\right)\,
P^{\ }_{\bs{Q}},
\end{equation}
of the resolution of the identity, we deduce that
\begin{equation}
P^{\ }_{\bs{Q}}=
P^{\ }_{\bs{Q}}\,P^{\ }_{\mathrm{e}}=
P^{\ }_{\mathrm{e}}\,P^{\ }_{\bs{Q}},
\qquad
\forall
\bs{Q}\neq\bs{Q}^{\ }_0.
\label{eq: PePQ=PQ}
\end{equation}

Now, we use the fact (see Ref.~\onlinecite{Karpus54})
that the position operator
$\bs{X}\equiv(X^{\ }_{i})$ can always be written as the
sum of two operators $\bs{T}\equiv(T^{\ }_{i})$ 
and $\bs{A}\equiv(A^{\ }_{i})$,
such that the former
shifts momentum by an infinitesimal amount in the
$i=1,\cdots,d$ direction and the latter does not shift 
the momentum,
i.e., $\bs{A}$ commutes with any $P^{\ }_{\bs{Q}}$,
\begin{equation}
\bs{X}=
\bs{T}
+ 
\bs{A}.
\label{eq: decomposition X}
\end{equation} 
It is of crucial importance to note that the decomposition%
~(\ref{eq: decomposition X}) is not unique, but basis dependent.%
~\cite{footnote: def X,Neupert12b}
Indeed, under those basis transformations of the single-particle Hilbert space
that are diagonal in momentum, i.e., those that commute with any $P^{\ }_{\bs{Q}}$,
the operator $\bs{A}$ transforms like an operator-valued gauge field.
Explicit representations of the operators $\bs{T}$  and $\bs{A}$ will be given
in Eqs.~\eqref{eq: operator T in Bloch basis} 
and~\eqref{eq: operator A in Bloch basis}, respectively.

We are now in position to do the following manipulations
for any $i=1,\cdots,d$. First, we do the decomposition
\begin{equation}
\begin{split}
P^{\ }_{\mathrm{e}}\,
X^{\ }_{i}\,
P^{\ }_{\mathrm{g}}
=&\,
P^{\ }_{\mathrm{e}}\,
\left(
T^{\ }_{i}
+
A^{\ }_{i}
\right)
P^{\ }_{\mathrm{g}}
\\
=&\,
P^{\ }_{\mathrm{e}}\,
\lim_{q\to 0}
\left(
\frac{
e^{\mathrm{i}\,q\,T^{\ }_{i}}-1
     }
     {
\mathrm{i}\,q
     }
\right)
P^{\ }_{\mathrm{g}}
+ 
P^{\ }_{\mathrm{e}}\,
A^{\ }_{i}\,
P^{\ }_{\mathrm{g}}.
\end{split}
\end{equation}
Second, we use the orthogonality 
$P^{\ }_{\mathrm{e}}\,P^{\ }_{\mathrm{g}}=0$
to dispose of the term $\mathrm{i}/q$
that would blow up in the limit 
of $q\to0$,
\begin{equation}
P^{\ }_{\mathrm{e}}\,
X^{\ }_{i}\,
P^{\ }_{\mathrm{g}}
=
\lim_{q\to 0}
P^{\ }_{\mathrm{e}}\,
\left(
\frac{
e^{\mathrm{i}\,q\,T^{\ }_{i}}
     }
     {
\mathrm{i}\,q
     }
\right)
P^{\ }_{\mathrm{g}}
+ 
P^{\ }_{\mathrm{e}}\,
A^{\ }_{i}\,
P^{\ }_{\mathrm{g}}.
\end{equation}
Third, we make use of the resolution of the identity
and the orthogonality from 
Eq.~(\ref{eq: resolution 1 in terms PQ and ortho PQ's})
together with Eqs.~(\ref{eq: PgPQ=0})
and~(\ref{eq: PePQ=PQ})
to infer that
\begin{equation}
P^{\ }_{\mathrm{e}}\,
X^{\ }_{i}\,
P^{\ }_{\mathrm{g}}
=
\lim_{q\to 0}
P^{\ }_{\bs{Q}^{\ }_0+q\,\bs{e}^{\ }_{i}}
\left(
\frac{
e^{\mathrm{i}\,q\,T^{\ }_{i}}
     }
     {
\mathrm{i}\,q
     }
\right)
P^{\ }_{\mathrm{g}}
+ 
P^{\ }_{\mathrm{e}}\,
A^{\ }_{i}\,
P^{\ }_{\mathrm{g}}\,
P^{\ }_{\bs{Q}^{\ }_0}.
\end{equation}
The first term on the right-hand side
connects two sectors of the Hilbert space
$\mathfrak{F}^{\ }_{N^{\ }_{\mathrm{e}}}$ 
with well-defined center-of-mass momenta 
differing by the momentum $q\,\bs{e}^{\ }_{i}$. 
The second term on the right-hand side annihilates 
any many-body state with $\bs{Q}\neq\bs{Q}^{\ }_0$.
Henceforth, the product
\begin{equation}
P^{\ }_{\mathrm{g}}\,
X^{\ }_{i}\,
P^{2 }_{\mathrm{e}}\,
X^{\ }_{j}\,
P^{\ }_{\mathrm{g}}
=
P^{\ }_{\mathrm{g}}\,
X^{\ }_{i}\,
P^{\ }_{\mathrm{e}}\,
X^{\ }_{j}\,
P^{\ }_{\mathrm{g}}
\end{equation}
on the right-hand side of Eq.~(\ref{eq: cond as trace over X})
becomes
\begin{equation}
P^{\ }_{\mathrm{g}}
X^{\ }_{i}
P^{\ }_{\mathrm{e}}
X^{\ }_{j}
P^{\ }_{\mathrm{g}}=
P^{\ }_{\mathrm{g}}\,
A^{\ }_{i}\,A^{\ }_{j}\,
P^{\ }_{\mathrm{g}}
-
P^{\ }_{\mathrm{g}}\,
A^{\ }_{i}\, 
P^{\ }_{\mathrm{g}}\,
A^{\ }_{j}\,
P^{\ }_{\mathrm{g}}
\label{eq: P X P X P}
\end{equation}
if $i\neq j$ (so that $P^{\ }_{q\,\bs{e}^{\ }_{i}}\neq P^{\ }_{q\,\bs{e}^{\ }_{j}}$)
and where we have assumed that we can freely
interchange the limit with the evaluation of the products. 

We now insert
Eq.~(\ref{eq: P X P X P})
into
Eq.~\eqref{eq: cond as trace over X},
\begin{equation}
\begin{split}
\bar{\sigma}^{\ }_{ij}
=&\,
-
\frac{\mathrm{i}}{2\pi\,V\,N^{\ }_{\mathrm{gs}}}\,
\mathrm{Tr}\,
\Big[
P^{\ }_{\mathrm{g}}[A^{\ }_{i},A^{\ }_{j} ]P^{\ }_{\mathrm{g}}
\\
&\,
-
(P^{\ }_{\mathrm{g}}\,A^{\ }_{i}\,P^{\ }_{\mathrm{g}})
(P^{\ }_{\mathrm{g}}\,A^{\ }_{j}\,P^{\ }_{\mathrm{g}})
+
(P^{\ }_{\mathrm{g}}\,A^{\ }_{j}\,P^{\ }_{\mathrm{g}})
(P^{\ }_{\mathrm{g}}\,A^{\ }_{i}\,P^{\ }_{\mathrm{g}})
\Big].
\end{split}
\end{equation}
The full trace $\mathrm{Tr}$ over the Hilbert space 
$\mathfrak{F}^{\ }_{N^{\ }_{\mathrm{e}}}$ 
is thus reduced to a trace over the ground state manifold. 
This is a \textit{finite sum}
since we have assumed that the ground state manifold is
a finite-dimensional vector space.
To dispose of the second contribution, we make
use of the cyclicity of the trace 
restricted to the ground state manifold. 
We are then left with the \textit{finite sum}
\begin{equation}
\begin{split}
\bar{\sigma}^{\ }_{ij}
=&\,
-
\frac{\mathrm{i}}{2\pi\,V\,N^{\ }_{\mathrm{gs}}}\,
\mathrm{Tr}\,
\left[
P^{\ }_{\mathrm{g}}[A^{\ }_{i},A^{\ }_{j} ]P^{\ }_{\mathrm{g}}
\right]
\\
=&\,
-\frac{\mathrm{i}}{2\pi\,V\,N^{\ }_{\mathrm{gs}}}\,
\sum_{n=1}^{N^{\ }_{\mathrm{gs}}}
\langle n|[A^{\ }_{i},A^{\ }_{j}]|n\rangle.
\end{split}
\label{eq: cond as commutator of A}
\end{equation}
Equation (\ref{eq: cond as commutator of A}) is the main
result of this paper. It is an algebraic counterpart 
to the many-body presentation
of the Hall conductance in terms of a many-body Berry phase
defined by twisting boundary conditions.%
~\cite{Niu85}

\section{
Bloch representation
        }
\label{sec: Bloch representation}

To proceed, we need to choose a basis of the
Hilbert space 
$\mathfrak{F}^{\ }_{N^{\ }_{\mathrm{e}}}$.
We choose the basis that follows from using the
Fock space $\mathfrak{F}$
spanned by the creation and annihilation
operators
$\chi^{\dag}_{a}(\bs{k})$
and
$\chi^{\ }_{a}(\bs{k})$,
respectively, whereby any such pair labeled
by the band index $a=1,\cdots,N$
and the Bloch momentum $\bs{k}$
corresponds to a Bloch state
$|\chi^{(a)}(\bs{k})\rangle$.
We choose the normalization conventions
\begin{equation}
\{\chi^{\ }_a(\bs{k}),\chi^{\dagger}_b(\bs{k}')\}=
\Omega\,\delta(\bs{k}-\bs{k}')\,\delta^{\ }_{a,b}\approx
\Omega\,V\delta^{\ }_{\bs{k},\bs{k}'}\,\delta^{\ }_{a,b}
\label{eq: fermion algebra}
\end{equation}
given the volume $V$ in position space set by the infrared cutoff
(the linear size $L\gg\mathfrak{a}$ say)
and the volume $\Omega$ in momentum space set by the ultraviolet cutoff
(the lattice spacing $\mathfrak{a}\ll L$ say).
In other words, we have assumed that 
$H^{\ }_{0}$
obeys the additive decomposition
\begin{subequations}
\begin{equation}
H^{\ }_{0}=
H^{\mathrm{Blo}}_{0}
+
H^{\mathrm{int}}_{0}.
\end{equation}
Here
\begin{equation}
H^{\mathrm{Blo}}_{0}=
\sum_{a=1}^{N}
\int\limits_{\Omega}
\frac{\mathrm{d}^{d}\bs{k}}{\Omega}\,
\varepsilon^{\ }_{a}(\bs{k})\,
\chi^{\dag}_{a}(\bs{k})\,\chi^{\ }_{a}(\bs{k})
\end{equation}
where, 
for any band index $a=1,\cdots,N$
and Bloch momentum $\bs{k}$,
the single-particle eigenvalue
$\varepsilon^{\ }_{a}(\bs{k})$
and the single-particle Bloch state
$|\chi^{(a)}(\bs{k})\rangle$
are the solution to the eigenvalue problem
\begin{equation}
\mathcal{H}^{\mathrm{Blo}}_{0}(\bs{k})\,
|\chi^{(a)}(\bs{k})\rangle=
\varepsilon^{\ }_{a}(\bs{k})\,
|\chi^{(a)}(\bs{k})\rangle.
\label{eq: def Bloch H0(k)}
\end{equation}
\end{subequations}
The interaction term 
$H^{\mathrm{int}}_{0}$
is of higher order than two in the number of
creation and annihilation operators.

The decomposition of the position operator
on the right-hand side of
Eq.~(\ref{eq: decomposition X})
can now be understood as follows.
Assume that the $N\times N$ Hermitian
matrix
$\mathcal{H}^{\mathrm{Blo}}_{0}(\bs{k})$
has been specified in the basis
$|\psi^{(\mathtt{a})}(\bs{k})\rangle$
where $\mathtt{a}=1,\cdots,N$.
We shall call this basis the orbital basis.
Diagonalization of 
$\mathcal{H}^{\mathrm{Blo}}_{0}(\bs{k})$
delivers the Bloch basis
$|\chi^{(a)}(\bs{k})\rangle$
where $a=1,\cdots,N$.
The unitary transformation that brings
the orbital to the Bloch basis
has the $N^{2}$ matrix elements
$u^{(a)}_{\mathtt{a}}(\bs{k})$
where $a,\mathtt{a}=1,\cdots,N$.
Any pair of columns or rows from this matrix
must be orthogonal, 
\begin{equation}
u^{(a) *}_{\mathtt{a}}(\bs{k})\,
u^{(b)}_{\mathtt{a}}(\bs{k})= 
\delta^{\ }_{a,b},
\qquad
u^{(a)}_{\mathtt{a}}(\bs{k})\,
u^{(a) *}_{\mathtt{b}}(\bs{k})= 
\delta^{\ }_{\mathtt{a},\mathtt{b}}.
\end{equation}
Here and in what follows, 
the summation convention over 
repeated band $a=1,\cdots,N$
or orbital index $\mathtt{a}=1,\cdots,N$
is implied.
For any spatial coordinate
$i=1,\cdots,d$, 
any pair 
$a,b=1,\cdots,N$
of bands, 
and any Bloch momentum $\bs{k}$
we define the 
non-Abelian Berry connection
\begin{equation}
A^{ab}_{i}(\bs{k}):=
-\mathrm{i} 
u^{(a) *}_{\mathtt{a}}(\bs{k})
\left(
\frac{
\partial
u^{(b)}_{\mathtt{a}}
     }
     {
\partial k^{\ }_{i}
     }
\right)(\bs{k}).
\label{eq: def Aabik} 
\end{equation}
The operators on the right-hand side of
Eq.~(\ref{eq: decomposition X}) 
are then represented by %
~\cite{Neupert12b}
\begin{equation}
\bs{T}=
\int\limits_{\Omega}
\frac{\mathrm{d}^{d}\bs{k}}{\Omega}\,
\chi^{\dag}_{a }(\bs{k})\,
\left(
\mathrm{i}
\frac{\partial\chi^{\   }_{a}}{\partial\bs{k}}
\right)(\bs{k})
\label{eq: operator T in Bloch basis}
\end{equation}
and
\begin{equation}
\bs{A}= 
\int\limits_{\Omega} 
\frac{\mathrm{d}^{d}\bs{k}}{\Omega} 
\chi^{\dag}_{a}(\bs{k})\,
\bs{A}^{ab}(\bs{k})\, 
\chi^{\ }_{b}(\bs{k}).
\label{eq: operator A in Bloch basis}
\end{equation}
Consequently, for any $i\neq j=1,\cdots,d$,
\begin{equation}
[A^{\ }_{i},A^{\ }_{j}]=
\int\limits_{\Omega} 
\frac{\mathrm{d}^{d}\bs{k}}{\Omega}
\chi^{\dag}_{a}(\bs{k})\,
[A^{\ }_{i}(\bs{k}),A^{\ }_{j}(\bs{k})]^{ab}\,
\chi^{\ }_{b}(\bs{k}).
\label{eq: rep [Ai,Aj]}
\end{equation}

Equation~(\ref{eq: rep [Ai,Aj]})
suggests that we define the $N^{2}$
dimensionless intensive numbers
\begin{subequations}
\begin{equation}
\bar{\mathrm{n}}^{ab}(\bs{k}):=
\frac{1}{\Omega\,V\,N^{\ }_{\mathrm{gs}}}
\sum_{n=1}^{N^{\ }_{\mathrm{gs}}}
\langle n | \chi^\dagger_a(\bs{k})\chi^{\ }_b(\bs{k}) | n\rangle.
\label{eq: def many-body nab(k)}
\end{equation}
With the normalization~(\ref{eq: fermion algebra}),
one verifies that
\begin{equation}
0\leq\bar{\mathrm{n}}^{aa}(\bs{k})\leq 1
\end{equation} 
\end{subequations}
for any band $a$ and any momentum $\bs{k}$.
Insertion of
Eq.~(\ref{eq: rep [Ai,Aj]}) 
into Eq.~(\ref{eq: cond as commutator of A}) 
yields
\begin{equation}
\bar{\sigma}^{\ }_{ij}=
-
\mathrm{i}
\int\limits_{\Omega} 
\frac{\mathrm{d}^{d}\bs{k}}{2\pi}\,
[A^{\ }_{i}(\bs{k}),A^{\ }_{j}(\bs{k})]^{ab}\, 
\bar{\mathrm{n}}^{ab}(\bs{k}).
\end{equation}

It remains to evaluate with the help of
Eq.~(\ref{eq: def Aabik})
the commutator 
$[A^{\ }_{i}(\bs{k}),A^{\ }_{j}(\bs{k})]^{ab}$
of the non-Abelian Berry connection.
It is
\begin{equation}
[A^{\ }_{i}(\bs{k}),A^{\ }_{j}(\bs{k})]^{ab}=
\mathrm{i}\,
\left(
\frac{\partial A^{ab}_{j}}{\partial k^{\ }_{i}}
\right)(\bs{k})
-
(i\leftrightarrow j).
\end{equation}
Hence, for any pair $i\neq j=1,\cdots,d$,
we conclude with the desired representation
\begin{equation}
\bar{\sigma}^{\ }_{ij}=
\int\limits_{\Omega}
\frac{\mathrm{d}^{d}\bs{k}}{2\pi}
\left[
\left(
\frac{\partial A^{ab}_{j}}{\partial k^{\ }_{i}}
\right)
(\bs{k})
-
(i\leftrightarrow j)
\right]
\bar{\mathrm{n}}^{ab}(\bs{k})
\label{eq: main rep bar sigma Bloch}
\end{equation}
of the Hall conductivity tensor averaged over the 
degenerate ground states.

\subsection{Noninteracting band insulator}
\label{subsec: Noninteracting band insulator}

For a noninteracting band insulator 
with the lowest 
$\widetilde{N}\leq N$ bands filled, 
we have $N^{\ }_{\mathrm{gs}}=1$
and
\begin{equation}
\bar{\mathrm{n}}^{ab}(\bs{k})=
\mathrm{n}^{ab}_{1}(\bs{k})=
\begin{cases}
\delta^{\ }_{a,b},
&
1\leq a,b\leq\widetilde{N},
\\
0,
&
\mathrm{otherwise}.
\end{cases}
\label{eq: occupation number band insulators}
\end{equation}
Due to the presence of the gap, 
the Bloch Hamiltonian%
~(\ref{eq: def Bloch H0(k)}) 
can be adiabatically deformed to
\begin{equation}
H^{\mathrm{Blo}}_{0}(\bs{k})
\longrightarrow 
1
-
2\,
\widetilde{P}(\bs{k}),
\label{eq: def deformed band insulator}
\end{equation}
where $\widetilde{P}(\bs{k})$ 
is the projector on the single-particle states 
in the lower bands. 
The right-hand side of Eq.~(\ref{eq: def deformed band insulator})
is invariant under any unitary transformation of the filled
bands, an element of the unitary group 
$U(\widetilde{N})$
which is a subgroup of the group of unitary transformations
$U(N)$ that mixes all $N$ bands.
The Hall conductivity~(\ref{eq: main rep bar sigma Bloch}) 
can be written 
in a form for which this symmetry manifests itself as a local
$U(\widetilde{N})$ gauge invariance,
\begin{equation}
\begin{split}
\bar{\sigma}^{\ }_{ij}=&\,
\int\limits_{\Omega}
\frac{\mathrm{d}^{d}\bs{k}}{2\pi}\, 
\widetilde{F}^{\tilde{a}\tilde{a}}_{ij}(\bs{k}).
\label{eq: band insulator unitary class}
\end{split}
\end{equation}
Here, the non-Abelian Berry curvature 
of the \textit{lower bands} is given by
\begin{equation}
\widetilde{F}^{\tilde{a}\tilde{b}}_{ij}(\bs{k}):=
\partial^{\ }_{k_{i}}A^{\tilde{a}\tilde{b}}_{j}(\bs{k})
+
\mathrm{i}\,
A^{\tilde{a}\tilde{c}}_{i}(\bs{k})\,
A^{\tilde{c}\tilde{b}}_{j}(\bs{k})\,
-(i\leftrightarrow j)
\end{equation}
and the indices with tildes
are summed only over the lower bands.
Note that
\begin{equation}
A^{\tilde{a}\tilde{c}}_{i}(\bs{k})\,
A^{\tilde{c}\tilde{a}}_{j}(\bs{k})
-
(i\leftrightarrow j)= 
A^{\tilde{a}\tilde{c}}_{j}(\bs{k})\,
A^{\tilde{c}\tilde{a}}_{i}(\bs{k})
-
(i\leftrightarrow j)
\end{equation}
vanishes if both $\tilde{a}$ and $\tilde{c}$ are summed over. 
This allows us to recast the expression for the Hall conductivity%
~(\ref{eq: main rep bar sigma Bloch}) 
in terms of the manifestly $U(\widetilde{N})$ 
gauge invariant Berry curvature%
~(\ref{eq: band insulator unitary class}).

\subsection{Noninteracting Fermi sea}
\label{subsec: Noninteracting Fermi sea}

Let us now consider a Fermi sea ground state for which
$N^{\ }_{\mathrm{gs}}=1$.
Even though our derivation relies on the existence 
of a gap from the outset, 
we can think of letting this gap go to zero 
at the end, as long as the ground state remains unique.
We shall work with a single partially occupied band $\tilde{a}=1$ 
for simplicity. With $\mathrm{FS}\subset\Omega$
denoting the Fermi sea 
\begin{equation}
\bar{\mathrm{n}}^{ab}(\bs{k})\equiv
\mathrm{n}^{ab}(\bs{k})=
\delta^{\ }_{a,1}
\times
\delta^{\ }_{b,1}
\times
\begin{cases}
1,&\bs{k}\in\mathrm{FS},
\\
0,&\bs{k}\in\Omega\setminus\mathrm{FS}.
\end{cases}
\label{eq: occupation number FS}
\end{equation}
The Hall conductivity~(\ref{eq: main rep bar sigma Bloch}) 
becomes
\begin{equation}
\begin{split}
\bar{\sigma}^{\ }_{ij}=&\,
\int\limits_{\mathrm{FS}\subset\Omega} 
\frac{\mathrm{d}^{d}\bs{k}}{2\pi}\, 
\widetilde{F}^{11}_{ij}(\bs{k}).
\end{split}
\end{equation}
This result agrees with 
the zero-temperature limit of the Hall conductivity 
for the anomalous Hall effect.%
~\cite{Ong06,AHE}

\subsection{
Interacting partially filled single Bloch band
           }
\label{subsec: Interacting partially filled single Bloch band}

Finally, we assume that the ground states $|n\rangle$ 
have nonvanishing amplitudes only with those Slater determinants 
that are made of single-particle Bloch states from the band
$\tilde{a}=1$.
The Hall conductivity~(\ref{eq: main rep bar sigma Bloch})
becomes
\begin{equation}
\begin{split}
\bar{\sigma}^{\ }_{ij}=&\,
\int\limits_{\Omega}
\frac{\mathrm{d}^{d}\bs{k}}{2\pi}\, 
\widetilde{F}^{11}_{ij}(\bs{k})\, 
\bar{\mathrm{n}}(\bs{k}),
\label{eq: single occupired Bloch band+interaction+gap}
\end{split}
\end{equation}
where $\bar{\mathrm{n}}(\bs{k})$ 
is the occupation number of the Bloch momentum $\bs{k}$
in that single band $\tilde{a}=1$ that contributes
to any one of the $N^{\ }_{\mathrm{gs}}$ ground states $|n\rangle$. 
The right-hand side
is invariant under any $U(1)$ gauge transformation of the 
$U(1)$ Berry connection $\bs{A}^{1}(\bs{k})$ that defines
the Berry curvature $\widetilde{F}^{11}_{ij}(\bs{k})$.

Equation (\ref{eq: single occupired Bloch band+interaction+gap})
sets bounds to the Hall conductivity 
that can arise in an interacting system from a partially occupied
single Bloch band. If we define the filling factor
\begin{equation}
\nu:= 
\int\limits_{\Omega}
\frac{\mathrm{d}^{d}\bs{k}}{\Omega}\,
\bar{\mathrm{n}}(\bs{k}),
\label{eq: def filling fraction}
\end{equation}
we conclude that
\begin{equation}
\frac{\Omega\,\nu}{2\pi}
\times
\inf_{\bs{k}\in\Omega\vphantom{\Omega^{T}}}\,
\widetilde{F}^{11}_{ij}(\bs{k})
\leq
\bar{\sigma}^{\ }_{ij}
\leq
\frac{\Omega\,\nu}{2\pi}
\times
\sup_{\bs{k}\in\Omega}
\widetilde{F}^{11}_{ij}(\bs{k})
\label{eq: bounds on sigma}
\end{equation}
for any pair $i,j=1,\cdots,d$.%
~\cite{footnote: bound on sigma}

When $d=2$, 
we deduce from Laughlin's flux insertion argument%
~\cite{Laughlin81} 
that 
\begin{equation}
\bar{\sigma}^{\ }_{ij} = 
\frac{p}{N^{\ }_{\mathrm{gs}}},
\label{eq: Hall conductivity as fraction}
\end{equation}
where $N^{\ }_{\mathrm{gs}}$ is the topological ground state degeneracy 
on the two-torus and $p$
is any integer that does not need to be co-prime with 
$N^{\ }_{\mathrm{gs}}$.
If we combine Eq.~\eqref{eq: Hall conductivity as fraction} 
with Eq.~(\ref{eq: bounds on sigma}), we conclude that
\begin{equation}
\frac{2\pi}{\Omega}
\left[
\sup_{\bs{k}\in\Omega}
F^{\ }_{ij}(\bs{k})
-
\inf_{\bs{k}\in\Omega\vphantom{\Omega^{T}}}
F^{\ }_{ij}(\bs{k})
\right]
\geq
\frac{1}{\nu}
\times
\frac{1}{N^{\ }_{\mathrm{gs}}}
\label{eq: inequality}
\end{equation}
is a necessary condition for the Hall conductivity to deviate from
\begin{equation}
\bar{\sigma}^{\ }_{12}= 
\nu\times C^{\ }_{12},
\qquad
C^{\ }_{12}:=
\int\limits_{\Omega}
\frac{\mathrm{d}^{2}\bs{k}}{2\pi}\, 
\widetilde{F}^{11}_{12}(\bs{k}),
\end{equation}
in two dimensions.  
Such a deviation
has been discussed for the FQHE in a periodic
potential~\cite{Kol93} and could in principle also appear in
two-dimensional fractional Chern insulators.%
~\cite{Shankar11}

We close by exploring another implication of 
Eq.~(\ref{eq: single occupired Bloch band+interaction+gap}) for 
fractional Chern insulators. 
It relates to the following question.
Can topologically ordered many-body 
states arise from a topologically trivial
single-particle band structure when interactions are added? 
Let us start by discussing two cases for which the answer
is negative, before turning to cases where the answer might be positive.

 First, according to 
Eq.~(\ref{eq: single occupired Bloch band+interaction+gap}),
if the single-particle Berry curvature vanishes everywhere in the
Brillouin zone,
$\widetilde{F}^{11}_{12}(\bs{k})=0$ for all $\bs{k}\in\Omega$, 
then the many-body Hall conductivity $\bar{\sigma}^{\,}_{12}$
has to vanish as well.
Second, if $\bar{\mathrm{n}}(\bs{k})$
is independent of $\bs{k}$ and if the band $\tilde{a}=1$ has
the vanishing Chern number $C^{\,}_{12}=0$, 
then $\bar{\sigma}^{\,}_{12}=0$. 

However, the condition that
$\bar{\mathrm{n}}(\bs{k})$ is constant throughout the Brillouin zone
of volume $\Omega$ is not required for a topologically ordered phase
of matter. When $\bar{\mathrm{n}}(\bs{k})$ varies in the Brillouin zone,
we can use the filling fraction $\nu$ 
defined by Eq.~(\ref{eq: def filling fraction})
and the inequality (\ref{eq: inequality})
to establish a necessary condition to be fulfilled by the variations
of the Berry curvature across the Brillouin zone of volume $\Omega$
for the many-body Hall conductivity $\bar{\sigma}^{\,}_{12}$ 
to acquire a nonvanishing value even though $C^{\,}_{12}=0$.

Even if the single-particle Berry curvature vanishes
or the necessary condition encoded by the inequality (\ref{eq: inequality})
is not fulfilled, a FQHE might still be stabilized by interactions
if the assumptions of 
Sec.~\ref{subsec: Interacting partially filled single Bloch band}
are relaxed. These assumptions are
that (i) only one isolated band is partially occupied and
(ii) discrete translational symmetry is not spontaneously broken.

If degrees of freedom from more than one band are available
or if discrete translational symmetry is spontaneously broken 
in such a way that the folding of the Brillouin zone results 
in new bands, interactions might then change the band structure 
from topologically trivial to non-trivial.
The former case is known to occur for Kramers degenerate bands.%
~\cite{Neupert12a}
For the latter case, we have in mind the scenario by
which a mean-field treatment of the interaction within
a single topologically trivial band breaks spontaneously
the discrete translational symmetry by reducing the Brillouin
zone from the volume $\Omega$ to the volume $\Omega^{\,}_{\mathrm{MF}}$.%
~\cite{scenario,Kalmeyer87,Laughlin88,Wen89}
In the process of folding the Brillouin zone from one with volume 
$\Omega$ to one with volume $\Omega^{\,}_{\mathrm{MF}}$, the original band
might split into several sub-bands (separated by energy gaps),
some of which carrying nonvanishing Chern numbers.
The residual interactions that have been ignored by this mean-field
treatment might then stabilize a FQHE characterized by
Eq.~(\ref{eq: single occupired Bloch band+interaction+gap}) 
provided $\Omega$ is substituted by $\Omega^{\,}_{\mathrm{MF}}$
and the original band $\tilde{a}$ is replaced by the relevant
sub-band. The ``spontaneous" formation of a fractional Chern insulator 
is thus allowed if more than one band is involved or discrete translational 
symmetry is spontaneously broken by the interaction.

\section*{Acknowledgments}

This work was supported in part by DOE Grant No.\ DEFG02-06ER46316
and by the Swiss National Science Foundation.


\begin{thebibliography}{99}
\bibitem{footnote: meaning translation inavriance}
By discrete translational invariance, we understand the space group
symmetry of some one-body periodic potential
that breaks Galilean invariance. This symmetry
is explicit in the Hamiltonian and not spontaneously broken
by the ground state manifold. This symmetry implies
that Bloch momenta are goood quantum numbers in the
noninteracting limit.
\bibitem{footnote: def FCI}
Fractional Chern insulators are strongly correlated phases of
matter for interacting fermions and bosons that have been found in
Bloch bands with vanishing or small bandwidth and nonvanishing Chern
number (Refs.%
~\onlinecite{Neupert11,Sheng11,Wang11a,Regnault11a,Bernevig11,Wang12a,Bernevig12a,Bernevig12b,Wang12b,Liu12}).
\bibitem{footnote: symm breaking}
If we were to work with a finite system instead,
there would be no notion of spontaneous symmetry breaking.
At the same time, the quasi-degenerate ground states 
could then have different center-of-mass momenta, depending,
for example, on the aspect ratio of the system. 
Whether these momenta converge to one single momentum 
or not as the thermodynamic limit is taken, dictates whether 
or not the translational symmetry will be broken spontaneously.
\bibitem{footnote: def X}
We refer the reader to 
Ref.~\onlinecite{Neupert12b}
for the precise definition of the many-body
position operator $\bs{X}$ and for the evaluation 
of the commutators of its components.
\bibitem{footnote: dimensions}
We call $\bs{J}$ a current because it couples linearly
to the vector gauge potential.
When $e=c=\hbar=1$, all dimensionfull quantities are measured
in units of energy:
$[\hbox{time}]=[\hbox{length}]=\hbox{energy}^{-1}$,
$[\bs{\mathcal{E}}]=\hbox{energy}^{+2}$,
$[\bs{\mathcal{A}}]=\hbox{energy}$,
so that the current
$\bs{J}$ is dimensionless,
$[\bs{J}]=\hbox{energy}^{0}$.
If $V$ denotes a finite volume of position space
in which the electrons
are constrained to move, then the current density
$\bs{j}\equiv\bs{J}/V$ has the dimensions 
$[\bs{j}]=\hbox{energy}^{+d}$.
\bibitem{footnote: bound on sigma}
Observe that the bound
\begin{equation}
|\bar{\sigma}^{\ }_{ij}|\leq\,
\int\limits_{\Omega}
\frac{\mathrm{d}^{d}\bs{k}}{2\pi}\, 
|\widetilde{F}^{11}_{ij}(\bs{k})|\, 
\bar{\mathrm{n}}(\bs{k})
\leq
\int\limits_{\Omega}
\frac{\mathrm{d}^{d}\bs{k}}{2\pi}\, 
|\widetilde{F}^{11}_{ij}(\bs{k})|
\nonumber
\end{equation}
also applies.
\end{thebibliography}
\end{document}